\documentclass[aps,preprint,prl,showpacs,floats,floats,superscriptaddress,floatfix]{revtex4-1}
\usepackage{bm}
\usepackage{verbatim}
\usepackage{graphicx}
\usepackage{graphics,epsfig}
\usepackage{theorem}
\usepackage{makeidx}
\usepackage{amsmath}
\usepackage{epic}
\usepackage{amscd}
\usepackage{xy}
\usepackage{amssymb}
\usepackage{epsfig}
\begin{document}
\newif\iffigs 
\figstrue
\iffigs \fi
\def\drawing #1 #2 #3 {
\begin{center}
\setlength{\unitlength}{1mm}
\begin{picture}(#1,#2)(0,0)
\put(0,0){\framebox(#1,#2){#3}}
\end{picture}
\end{center} }

\newcommand{\pkg}{\mathrm{P}_{\!\!\! _{K_{\scriptscriptstyle \mathrm{G}}}}}
\newcommand{\ul}{{\bm u}_{\scriptscriptstyle \mathrm{L}}}
\newcommand{\deltal}{{\delta}_{\scriptscriptstyle \mathrm{L}}}
\newcommand{\ue}{\mathrm{e}}
\newcommand{\ui}{\mathrm{i}\,}
\newcommand{\kg}{{k_{\scriptscriptstyle \mathrm{G}}}}
\def\v{\bm v}
\def\x{\bm x}
\def\k{\bm k}
\def\ds{\displaystyle}

\title{Transition from dissipative to conservative dynamics in equations of hydrodynamics}
\author{Debarghya Banerjee}
\email{debarghya@physics.iisc.ernet.in}
\affiliation{Centre for Condensed Matter Theory,
Department of Physics, Indian Institute of Science,
Bangalore 560012, India}
\author{Samriddhi Sankar Ray}
\email{samriddhisankarray@gmail.com}
\affiliation{International Centre for Theoretical Sciences, Tata Institute of Fundamental Research, Bangalore 560012, India}

\begin{abstract}

We show, by using direct numerical simulations and theory, how, by increasing
the order of dissipativity ($\alpha$) in equations of hydrodynamics, there is a
transition from a dissipative to a conservative system. This remarkable result,
already conjectured for the asymptotic case $\alpha \to \infty$  [U. Frisch et
al., Phys. Rev. Lett.  {\bf 101}, 144501 (2008)], is now shown to be true for
any large, but finite, value of $\alpha$ greater than a crossover value
$\alpha_{\rm crossover}$. We thus provide a self-consistent picture of how
dissipative systems, under certain conditions, start behaving like conservative
systems and hence elucidate the subtle connection between equilibrium
statistical mechanics and out-of-equilibrium turbulent flows. 

\end{abstract}

\keywords{Bottlenecks, Thermalisation, Hyperviscosity }

\date{\today}
\pacs{47.27.Gs, 47.10.ad}

\maketitle

Since the pioneering work of E. Hopf~\cite{hopf} and T. D. Lee~\cite{lee}, over
60 years ago, physicists have tried to understand the strongly
out-of-equilibrium, dissipative turbulent flows by using the tools of classical
equilibrium statistical mechanics. What makes such attempts particularly
difficult is that although, from a microscopic point of view, fluid motion can
be modelled via a Hamiltonian formulation, with statistically steady states
governed by an invariant Gibbs measure, a self-consistent macroscopic approach
inevitably leads to a dissipative hydrodynamical description with an
irreversible energy loss through heat dissipation at the molecular level.  In
the last few years, however, significant work has gone into our understanding
of the interplay between equilibrium statistical mechanics and turbulent flows
~\cite{brachet05,ray11,frisch12,majda00,lvov02}.   In particular, the
thermalised solutions to the Galerkin-truncated equations of hydrodynamics,
such as the three- (3D) or two-dimensional (2D) Euler~\cite{brachet05,GK09}, Gross-Pitaevskii~\cite{GP} and
magnetohydrodynamic~\cite{GKMHD11} equations and the one-dimensional (1D) Burgers equation,
have been studied extensively by several authors
\cite{majda00,ray11}.  For example, it is possible to obtain a
conservative dynamical system, which obeys Gibbsian statistical mechanics, for
hydrodynamical equations of an ideal fluid where only a finite number of
Fourier modes are retained by using the method of Galerkin
truncation~\cite{brachet05,ray11}.  Indeed since the first prediction of such
thermalised states ~\cite{kraich89}, its existence was shown by Cichowlas et
al.~\cite{brachet05}, for the incompressible, truncated 3D Euler equations, and
the explanation of how thermalisation sets in such systems was given by Ray et
al.~\cite{ray11} through the phenomenon of {\it tygers}.

Much of the work discussed above for thermalised states was done for
finite-dimensional, conservative systems obeying a Liouville theorem.
Therefore it is important to ask if there are  connections between such states
and dissipative, turbulent flows described by viscous Navier--Stokes-like
equations. A partial answer was given by Frisch et al.~\cite{frisch08}, where
the energy spectrum bottleneck, a bump in the spectrum between the inertial and
dissipation ranges, in solutions of the incompressible 3D Navier-Stokes and the
compressible 1D Burgers equation was attributed to an aborted thermalisation.
By using direct numerical simulations (DNSs) and
Eddy-Damped-Quasi-Normal-Markovian (EDQNM) calculations~\cite{edqnm}, it was
shown that if we replace the usual viscous operator $\nu \nabla^2 {\bf u}$ by
the hyperviscous operator $-\nu (-\nabla^2)^{\alpha} {\bf u}$, where $\nu$ is
the coefficient of viscosity, $\alpha$ is the order of hyperviscosity
(dissipativity), and $\bf u$ the velocity field, the bottleneck becomes
stronger with increasing $\alpha$.  The authors observed
that for extremely large values of $\alpha \ge 500$, the bottleneck is 
due to partial thermalisation observed in ~\cite{brachet05}. This
lead to the intriguing conjecture that the usual bottlenecks, seen in solutions
of the Navier--Stokes equation ~\cite{dobler03,she93,kaneda0309,donzis10} and
in experiments~\cite{dobler03,pak91} is possibly because of aborted
thermalisation.

The large $\alpha$ limit~\cite{frisch08} is extremely important from the point
of view of our understanding of hydrodynamical equations. However in most DNSs,
which seek to increase the effective inertial range via hyperviscosity, much
smaller values of $\alpha \le 16$ are typically used, which, nevertheless
produce significant bottlenecks.  For the ordinary Navier--Stokes equation
($\alpha = 1$), a theoretical understanding of the bottleneck was proposed in
~\cite{falko94}.  Recently, a more complete explanation of this effect was
given in ~\cite{frisch13} where it was shown that this bottleneck has its
origins in oscillations in the velocity correlation function. This mechanism
is, {\it apparently}, very different to the aborted thermalisation for large
$\alpha$ proposed in ~\cite{frisch08}.

Can this {\it apparent paradox}, when going from small to large values of
$\alpha$, be resolved? In this paper we show how, by increasing the order of
dissipativity ($\alpha$), we can crossover from one regime~\cite{frisch08} to
another~\cite{frisch13} and thus resolve the paradox. More importantly, as we
explain below, our work shows how the tuning of a single parameter $\alpha$ can
change a dissipative system to one which displays features of a conservative,
Hamiltonian system, leading to a thermalised state. This remarkable result was
already conjectured in ~\cite{frisch08} for the asymptotic case $\alpha \to
\infty$; in this paper we show, by using both DNSs and theory, that this is
already the case for a large, but finite, value of $\alpha$.  We thus provide a
self-consistent picture of how dissipative systems can start behaving like
conservative systems and thus elucidate the subtle connection between
equilibrium statistical mechanics and out-of-equilibrium turbulent flows. 

\begin{figure}
\includegraphics[width=0.8\columnwidth]{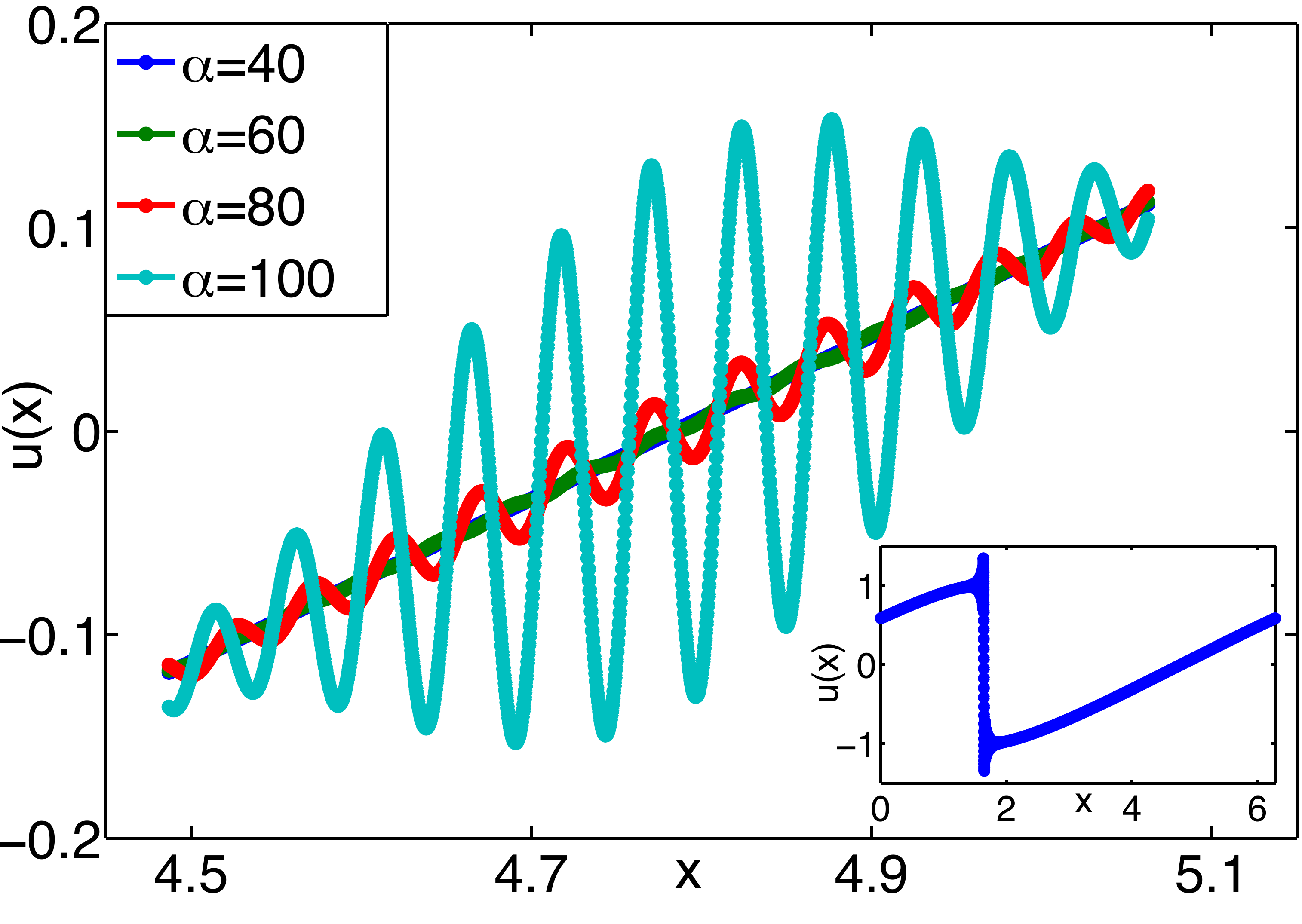}
\caption{(color online) Solutions of the HBE, zoomed around $x_{\rm s}$, for various values 
of $\alpha$ (see legend) at $t = 1.5$ showing oscillations at $x_{\rm s}$ 
with increasing amplitude as $\alpha$ increases (see text). 
Inset : Solution of the HBE with no oscillations at $x_{\rm s}$ for small $\alpha = 20$.}
\label{fig:osc}
\end{figure}

The Burgers equation has had a long history of being a testing ground 
for such ideas related to fluid dynamics~\cite{burgreviews}, and, more 
recently the chaotic behaviour in conservative systems~\cite{ray11}. 
Therefore, we begin, with the 1D, unforced, hyperviscous Burgers equation (HBE) : 
\begin{equation}
\frac{\partial u}{\partial t} + \frac{1}{2} \frac{\partial u^2}{\partial x} = 
-\nu {\left(-\frac{1}{k_d^2}\frac{\partial^2}{\partial x^2}\right )}^{ \alpha} u, 
\label{eq:hypburg}
\end{equation}
where, $u$ is the velocity field, $x$ and $t$, the space and time variables,
respectively; $\nu$ is the coefficient of kinematic hyperviscosity, and $k_d$ a reference
wavenumber.  In the limit of vanishing viscosity $\nu \to 0$, 
with $\alpha \ge 2$, the solution to Eq. (\ref{eq:hypburg}) develop oscillations 
in the boundary layer around the shock; these oscillations result in a 
bottleneck in the Fourier space energy spectrum ~\cite{frisch13}. 
These oscillations -- which have been studied by using 
boundary-layer-expansion techniques~\cite{frisch13} --  are localised in the neighbourhood of the shock and 
decay exponentially as one moves away. The wavelength $\lambda^{\rm th}_\alpha$ and the decay rate $K_{\alpha}^{\rm th}$
of these oscillations are given by 
\begin{eqnarray}
\lambda^{\rm th}_\alpha &=& 2\pi \nu^{\beta} k_d^{-2 \alpha \beta}\left [2^\beta \sin[(2n_\star+1)\beta \pi)]\right ]^{-1}
\label{lambdatheory} \\
K_{\alpha}^{\rm th} &=& 2^{\beta} \nu^{-\beta} k_d^{2 \alpha \beta} \cos[(2n_*+1)\beta \pi];
\label{eq:efolding}
\end{eqnarray}
where, $\beta = \frac{1}{2\alpha -1}$ and $n_*$ is an integer, $0 \le n_* \le 2\alpha - 2$, whose
value is obtained via linearisation and boundary layer analysis ~\cite{frisch13}.

For extremely large values of $\alpha \ge 500$, the solution of Eq. (\ref{eq:hypburg}) 
start thermalising~\cite{frisch08} and, at very large times, becomes 
indistinguishable from the solution $v(x,t)$ of the associated Galerkin-truncated (inviscid), {\it conservative},  
Burgers equation~\cite{ray11} : $\frac{\partial v}{\partial t} + \pkg \frac{1}{2} \frac{\partial v^2}{\partial x} = 0$, 
where the Galerkin projector $\pkg$ is a low-pass filter which sets to zero all Fourier components with wavenumbers $|k|>\kg$

At this stage, it behooves us to ask the question what happens for intermediate values of 
$\alpha$? And, furthermore, is there a single mechanism which can self-consistently describe 
the transition from a turbulence regime to a thermalised state in equations of
hydrodynamics?

\begin{figure}
\includegraphics[width=0.8\columnwidth]{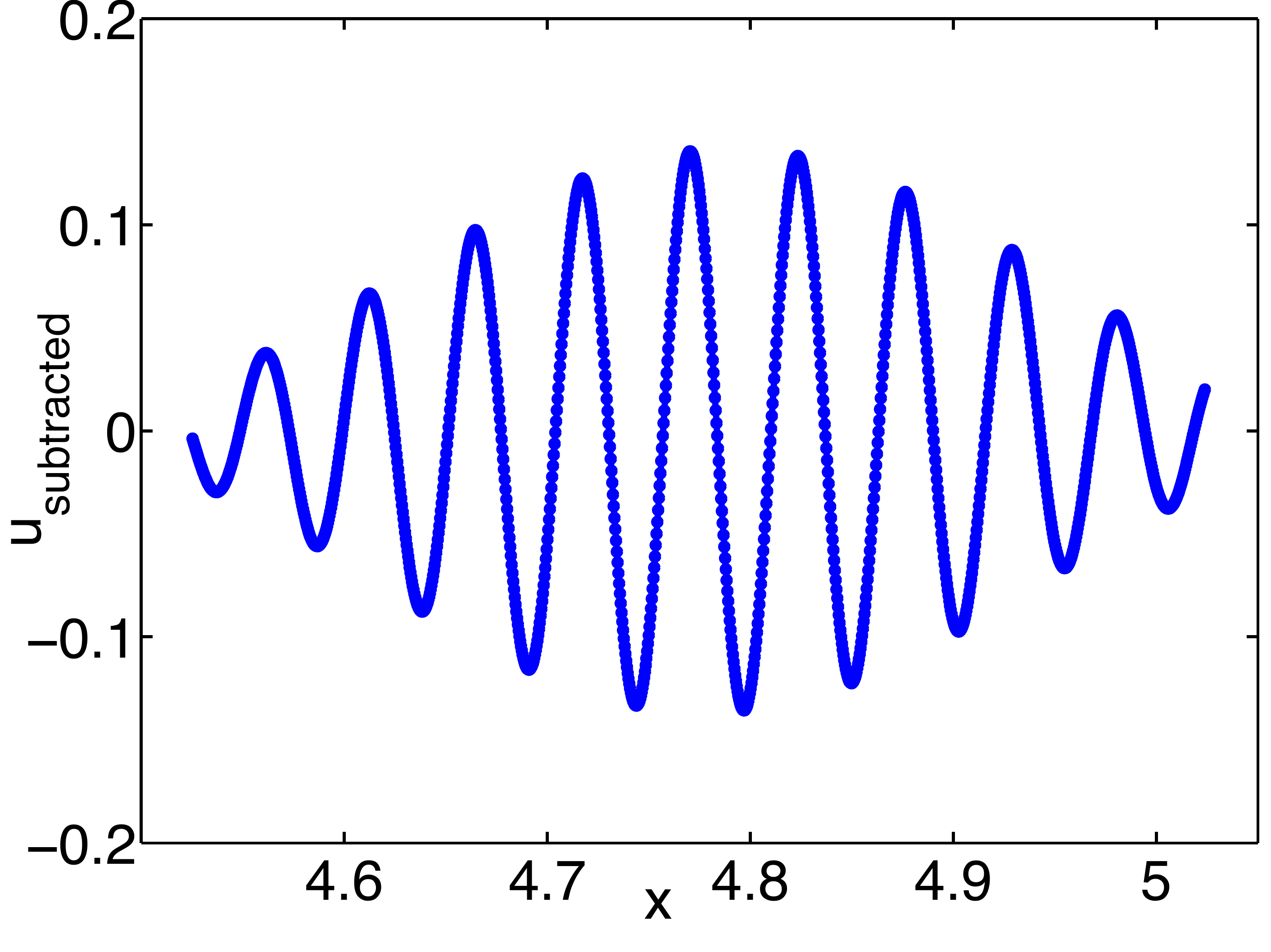}
\caption{(color online) The solution of the HBE for $\alpha = 100$, 
with the solution for the ordinary Burgers subtracted out, zoomed around $x_{\rm s}$. A 
clear symmetric bulge, similar to those seen in inviscid, conservative truncated system ~\cite{ray11}, is seen.}
\label{fig:bulge}
\end{figure}

The onset of thermalisation, in inviscid, finite-dimensional systems of the
Euler or the Burgers equation, is due to the birth of structures called tygers.
These are caused~\cite{ray11} by the motion of 
fluid particles interacting resonantly with the waves generated, because of truncation, 
by small-scale features, such as shocks. The special points
in physical space where tygers appear are the so-called {\it stagnation
points} $x_{\rm s}$, which, in the case of the Galerkin-truncated Burgers
equation, are points which have the same velocity as the shock(s) and a
positive local gradient. Is there another way, apart from truncation waves in
inviscid systems, for waves to be generated at the stagnation points in a fluid
for similar resonant interactions leading to an onset of thermalisation? We will 
show that for $\alpha$ greater than a crossover value $\alpha_{\rm crossover}$, 
a significant fraction of the oscillations, governed by Eqs. (2) and (3), which start
from the boundary layer near the shock, must reach  $x_{\rm s}$
and trigger tyger-like structures leading to thermalisation.

In order to answer these questions we first perform pseudo-spectral DNSs of Eq. (\ref{eq:hypburg})
on a $2\pi$ periodic line, with a
second-order Runge--Kutta scheme for time-integration. We
use a time step $\delta t = 10^{-4}$, the number of collocation points
$N=16384$, $\nu = 10^{-20}$ and $k_d=100$. Crucially, we use  $2 \le \alpha \le 500$
to study this intriguing transition from dissipative
dynamics to conservative, thermalised states. 
Our initial condition $u_0(x) = \sin (x + 1.5)$ leads to
$x_{\rm s} = 2\pi - 1.5 \approx 4.8$ and, in the absence of viscosity, 
shock formation at time $t_* = 1.0$. 

\begin{figure}
\includegraphics[width=0.8\columnwidth]{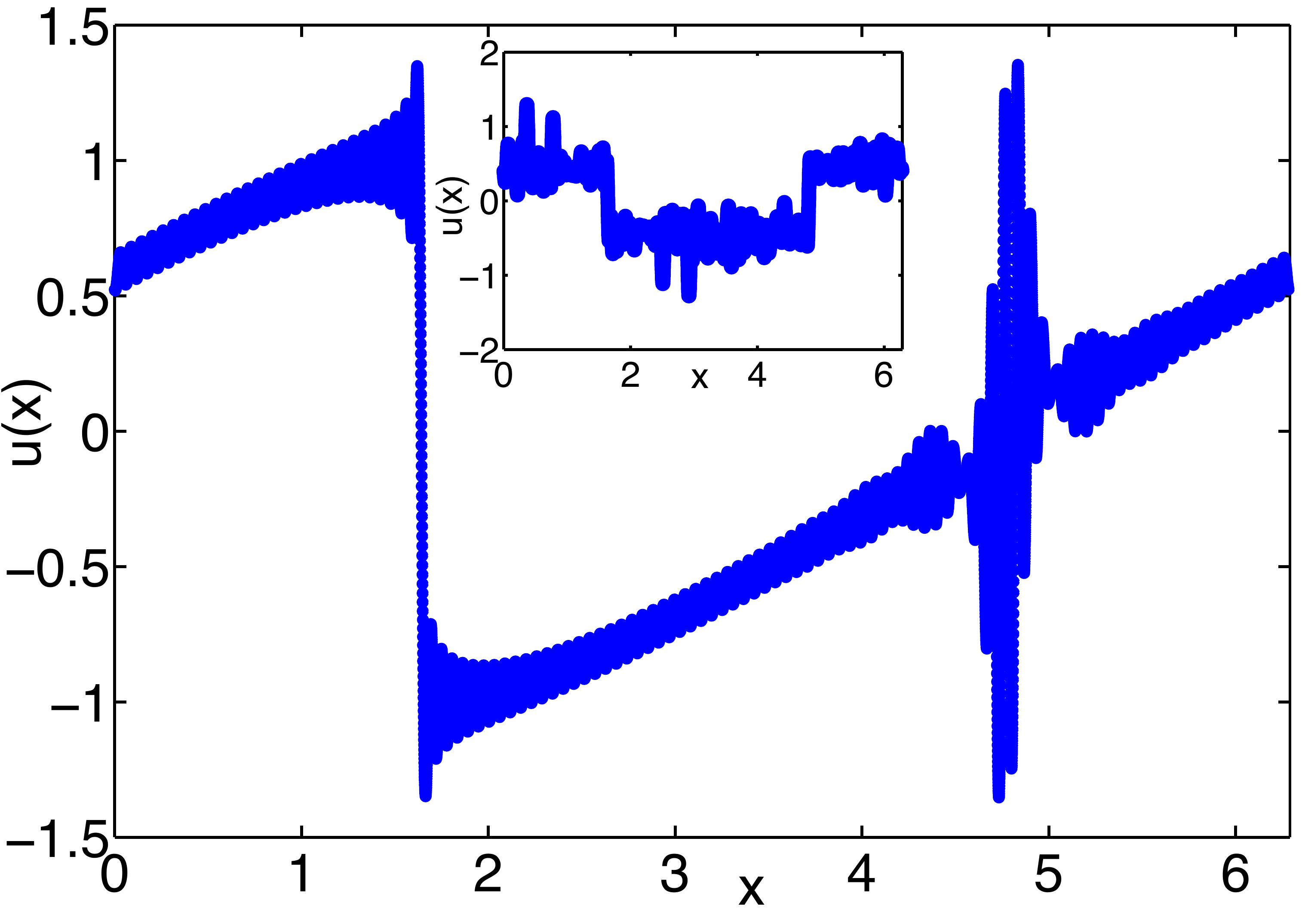}
\caption{(color online) Solution of the HBE, for 
$\alpha=250$ at $t=1.5$, showing the presence of a {\it tyger} ~\cite{ray11} 
at $x_{\rm s}$. Inset : $u(x)$ for $\alpha = 250$ at a later time 
($t=5.0$) confirming that for $\alpha \gtrsim \alpha_{\rm crossover}$ the system eventually thermalises.}
\label{fig:thermal}
\end{figure}

We begin our simulations from $\alpha = 2$ and observe~\cite{frisch13} that
with increasing $\alpha$, oscillations in a thin layer around the shock become
pronounced with a  related bottleneck in the energy spectrum. However, near the
stagnation point $x_{\rm s}$, no oscillations are seen for $\alpha \lesssim
40$.  This is clearly seen in a plot of $u(x)$ versus $x$, as shown in the
inset of Fig.~\ref{fig:osc}, at $t = 1.5$ for $\alpha = 20$.  However, as
$\alpha$ increases, finite, but small, oscillations start to reach $x_{\rm s}$
from the boundary layer around the shock. This is shown in Fig.~\ref{fig:osc}
for values of $\alpha = 40,60,80,$ and 100.  Furthermore, for values of $\alpha
\gtrsim 80$, a distinct bulge, reminiscent of the tygers found in solutions of
the Galerkin truncated equation ~\cite{ray11}, is clearly seen at $x_{\rm s}$.
This, then, is the first evidence of what triggers thermalisation in a
dissipative system and whose dramatic consequences were studied in
Ref.~\cite{frisch08} for the special case of $\alpha \to \infty$.

How similar is this bulge, in the dissipative HBE, at $x_{\rm s}$ for $\alpha
\gtrsim 80$, to that seen at $t_*$ for the Hamiltonian system of the Galerkin
truncated Burgers equation? In order to answer this question, it is useful to
examine the bulge, via $u_{\rm subtracted} = u - U$, where $U$ is the
(non-oscillatory) solution of the inviscid Burgers equation. In
Fig.~\ref{fig:bulge} we show this subtracted bulge for $\alpha = 100$ and find
that this bulge has the same shape as tygers~\cite{ray11} 
and is also symmetric around $x_{\rm s}$. The wavelength of these oscillations, as is
expected from a resonance build-up argument, is the same as the wavelength of
the oscillations emanating from the boundary layer around the shock~\eqref{lambdatheory}. 
A significant difference between the bulge observed
for moderate values of $\alpha$ (Figs.~(\ref{fig:bulge}) and (\ref{fig:osc})), 
and that of a tyger~\cite{ray11}, is its large width and the surprisingly small
number of oscillations inside it. In the truncated system, the bulge
width is proportional to $\kg^{-1/3}$ and the wavelength of the oscillations
proportional to $1/\kg$; this yields the number of oscillations in the bulge to
be proportional to $\kg^{2/3}$. In the present problem, the width of the bulge
is explained as follows : At time $t$, such that $\tau = t - t_*$, and when the
bulge is still symmetric around $x_{\rm s}$, resonant interactions are confined
to particles such that $\tau\Delta u \equiv \tau|u - u_{\rm s}| \lesssim
\lambda^{\rm th}_\alpha$, where $u_{\rm s}$ is the velocity of the shock. We
have chosen $t$ such that $\tau \sim 1$, leading to $\Delta u \sim \lambda^{\rm
th}_\alpha$.  Given that around $x_{\rm s}$ the velocity is proportional to
$x$, this yields a bulge width $\sim \lambda^{\rm th}_\alpha$ with a few
oscillations inside.

\begin{figure}
\includegraphics[width=0.8\columnwidth]{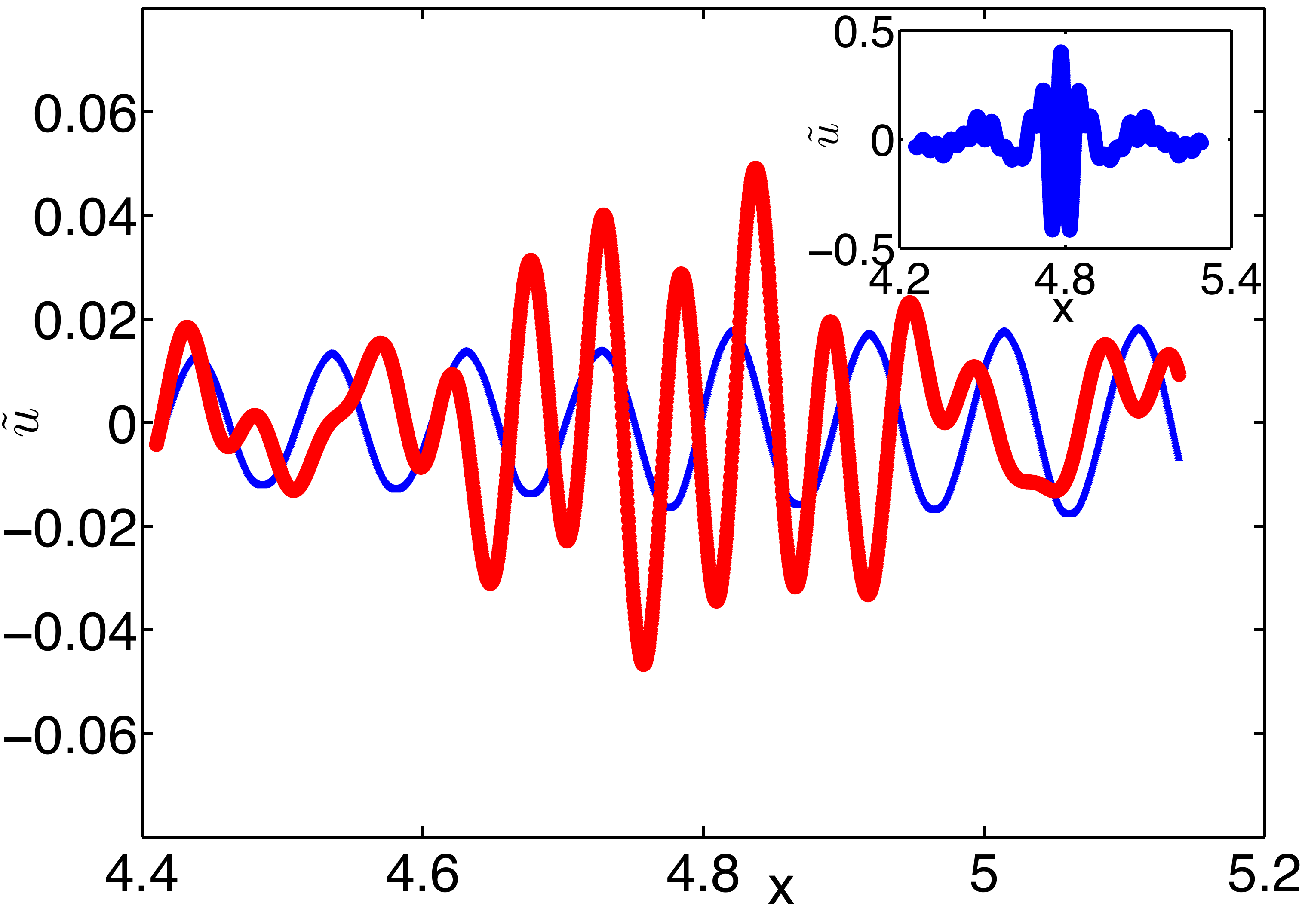}
\caption{(color online) The solution $\tilde u$ of the linearised equation \eqref{lineq} 
with initial conditions I1 at time $t = 2.0$ (blue curve) and $t=2.5$ (red curve) for 
$\alpha = 100$ (see text). Inset : $\tilde u$ with initial conditions I2 at time $t=10.0$.}
\label{fig:sol_lin}
\end{figure}

In the case of the inviscid, Galerkin truncated Burgers equation, the early
bulge (tygers) become asymmetric in time, leading to an eventual collapse
and thermalised states. In the present dissipative problem, 
although for reasonably small values of $\alpha$, a bulge is
guaranteed to form at $x_{\rm s}$, its eventual dynamics -- and indeed whether
the system actually thermalises -- depends on the interplay between the local
dissipation around $x_{\rm s}$, the fraction of oscillations reaching $x_{\rm
s}$ from the boundary layer, and the effect of the nonlinearity. For smaller
values of $\alpha$, when the dissipation is strong and the amplitude of
oscillations is small, this bulge at large times, remains stationary in time 
without ever collapsing and leading to a complete thermalisation.  However, as
$\alpha$ increases, the amplitude of oscillations reaching the stagnation point
is significant : Consequently for values of $\alpha$ higher than
a threshold $\alpha_{\rm crossover}$, the local dissipation can no longer
compensate for the resonant pile up at $x_{\rm s}$ leading to the emergence of
thermalised states in a manner exactly similar to that of the inviscid
truncated systems.  Heuristically, an estimate of $\alpha_{\rm crossover}$ can
be obtained as follows : The fraction of amplitude at the boundary layer that
reaches $x_{\rm s}$ is given by $e^{-K_{\alpha}^{th} \pi}$.  We assume that a
significant level of oscillations is present at $x_{\rm s}$ when at least a
fraction 1/e of the oscillations produced near the shock reaches $x_{\rm
s}$, i.e., $K_{\alpha}^{th} \pi = 1$. By using \eqref{eq:efolding}, and in the
limit of large $\alpha$, we obtain $\alpha_{\rm crossover} = \frac{1}{2}(1+50
\pi^2) \simeq 230$.  

We now examine the accuracy of our estimate of $\alpha_{\rm crossover}$ through
detailed simulations with increasing values of $\alpha$. As we increase
$\alpha$, our simulations show that the bulge at the $x_{\rm s}$ reaches a
stationary state without collapsing. However at around $\alpha \gtrsim 220$ we
observe that the bulge which forms, due to resonance, at $x_{\rm s}$, collapses
in a finite time and then the system thermalises in a manner reminiscent to the
dynamics of the Galerkin truncated Burgers equation~\cite{ray11}.  This is best
seen in Fig.  (\ref{fig:thermal}) where we show the solution of the HBE for
$\alpha = 250$ at time $t = 1.5$. We note that, just as in the inviscid,
Galerkin truncated Burgers equation~\cite{ray11}, the bulge at the resonance
point becomes very large, assymetric and non-monochromatic with secondary
structures on either side of it. This is exactly similar to the onset of
thermalisation in conservative systems~\cite{ray11}. Indeed at larger times the
solution completely thermalises (inset of Fig.  (\ref{fig:thermal}), at $t =
5.0$).  Our simulations illustrate quite clearly that (a) the heuristic
estimate of $\alpha_{\rm crossover}$ is correct and, more importantly, (b)
dissipative systems, such as the HBE, can thermalise at finite values of the
order of dissipativity in a manner similar to that of conserved, truncated
systems. The fact that dissipative systems can start to mimic truncated,
Hamiltonian system through the tuning of a single parameter ($\alpha$) is
a striking result and resolves a long standing paradox in the area of
turbulence and statistical mechanics. It is not hard to conjecture that this
cross-over should be possible for $\alpha \to \infty$~\cite{frisch08}; however,
remarkably, we now show that the onset to thermalisation actually occurs at a
finite value $\alpha_{\rm crossover}$.

Let us finally address the question of whether this phenomenon can be captured  
within a systematic theoretical framework. Rewriting Eq.~\eqref{eq:hypburg} in terms 
of the solution $U$ of the inviscid Burgers equation and the discrepancy $\tilde u \equiv u - U$, and using  
$\frac{\partial U}{\partial t} + \frac{1}{2} \frac{\partial U^2 }{\partial x} = 0$, we obtain, 
\begin{equation}
\frac{\partial \tilde u}{\partial t} + \frac{\partial}{\partial x} (U \tilde u) + \frac{1}{2} \frac{\partial {\tilde u}^2}{\partial x}  = -\nu \left(- \frac{1}{k_d^2}\frac{\partial^2 }{\partial x^2} \right)^{\alpha} \left (\tilde u + U\right).
\label{rewrite}
\end{equation}
At times close to $t_*$, and away from the shock, the discrepancy  between the solution of the  
HBE~\eqref{eq:hypburg} and the solution of the
inviscid Burgers equation is small ($\tilde u/U \ll 1$); hence we can drop the 
quadratic term. Next, we note that $U$ is linear 
in the spatial variable $x$ away from the shock which implies that higher derivatives of
$U$ vanish around $x_{\rm s}$. By using these two approximations, we finally obtain the following, analytically more 
tractable, linear equation :
\begin{equation}
\frac{\partial \tilde u}{\partial t} + \frac{\partial}{\partial x} (U \tilde u) = -\nu \left(- \frac{1}{k_d^2}\frac{\partial^2 }{\partial x^2} \right)^{\alpha} \tilde u.
\label{lineq}
\end{equation}

We first validate our linear theory by numerically solving \eqref{lineq} for
$\tilde u$ with a further approximation that $U$ is the solution at $t_*$ of
the inviscid Burgers equation, with the initial condition $\sin (x + 1.5)$.  We
choose two kinds of initial conditions $\tilde u_0 = \tilde u (t=0) $ : (I1)
$\tilde u_0$ is a low amplitude sinusoidal function with a wavenumber equal to
10; and (I2) $\tilde u_0 = e^{-K_{\alpha}^{th} |x-x_{\rm shock}|} \sin{\frac{2
\pi (x-x_{\rm shock})}{\lambda_{\alpha}^{th}}}$, where $x_{\rm shock}$ is the
position of the shock. Our numerical integration of Eq.~\eqref{lineq} for both
classes of initial conditions yield similar results as illustrated in Fig.
~\ref{fig:sol_lin} where we present a representative plot of $\tilde u$, solved
for Eq.~\eqref{lineq}, at time $t = 2$ (blue curve), and, $t=2.5$ (red curve)
for $\alpha = 100$ by using the initial conditions I2; the inset shows the
solution of Eq.~\eqref{lineq} for initial conditions I1 at time $t=10.0$. A
symmetric bulge at the stagnation point, just like in the solutions
Eq.~\eqref{eq:hypburg} for large $\alpha$ is clearly seen.  The essential
features of the bulge, i.e., its locality and the fact that it forms at the
stagnation point is reproduced by our linear model. Having established the
validity of the linear model to predict the location and the nature of the
bulge, we can now solve Eq.~\eqref{lineq} by various standard analytical means
such as by using the method of separation of variables or through a Fourier
transform of Eq.~\eqref{lineq}, to obtain solutions of Eq.~\eqref{lineq} (upto
constants) which show the existence of symmetric bulges at $x_{\rm s}$ which
decay on either side of the stagnation point.  We should note in passing, that
although the linear model predicts well the early stages of the
formation of the bulge at $x_{\rm s}$, our extensive simulations of the linear
model, for various large values of $\alpha$, not surprisingly, fails capture
the collapse of the bulge and eventual thermalisation~\cite{ray11}.  A
plausible conjecture for this is that the non-linearity, however weak, is
responsible for the stretching of the bulge and generating an associated
Reynolds stress which must be present to make the symmetric bulge collapse and
trigger complete thermalisation.

For the past many decades, a vexing and open question in the areas of
turbulence and statistical mechanics is how meaningful are thermalised states
in such problems.  In this paper we answer this question via detailed numerical
simulations and linear models. Our results show that just as in the case of
Hamiltonian systems of the Galerkin-truncated equation, where monochromatic
truncation waves can reach $x_{\rm s}$, leading to an accumulation, via
resonance, and eventual thermalisation, similarly, for dissipative systems such
as the HBE, for moderately large $\alpha$, monochromatic boundary layer
oscillations reach and accumulate, via the same resonant effect, at $x_{\rm
s}$. These bulges are the seeds of an eventual thermalised regime and for
$\alpha \gtrsim \alpha_{\rm crossover}$ the dissipative system does thermalise
at large times in a manner similar to the inviscid truncated system. Our work
thus connects the apparently disconnected worlds of conservative and
dissipative systems. Although we have confined ourselves to the one-dimensional
Burgers equation, the central result obtained in this paper should be valid in
the multidimensional Navier--Stokes equation for the reasons outlined in
Refs.~\cite{frisch08, ray11, frisch13}. A detailed study of this is beyond the
scope of the present paper and is left for the future.

\begin{acknowledgements}
We are grateful to M.-E. Brachet, U. Frisch, T. Matsumoto, S. Nazarenko, and R.
Pandit for many useful discussions.  We thank CSIR, UGC, DAE, and DST (India)
for support, and SERC (IISc) for computational resources.  We acknowledge the
support of the Indo--French Center for Applied Mathematics (IFCAM).  SSR
acknowledges support from EADS Corporate Foundation Chair awarded to ICTS-TIFR
and TIFR-CAM.
\end{acknowledgements}


\end{document}